\documentclass[referee]{aa} 
\begin{document}
\title{Discovery of A New Faint Radio SNR G108.2-0.6}
\author{W.W. Tian \inst{1,2}
\and
      D.A. Leahy \inst{2}
\and
   T.J. Foster \inst{3}}
\authorrunning{Tian, Leahy \& Foster}
\offprints{Wenwu Tian}
\institute{National Astronomical Observatories, CAS, Beijing 100012, China \\
\and
Department of Physics \& Astronomy, University of Calgary, Calgary, Alberta T2N 1N4, Canada\\
\and
Department of Physics \& Astronomy, Brandon University, Brandon, Manitoba, R7A 6A9 Canada}
 
\date{received Dec. 18, accepted xx, 2006} 
\abstract{A new faint and large shell-type radio Supernova Remnant (SNR) 
G108.2$-$0.6 has been discovered in the Canadian Galactic Plane Survey (CGPS). 
The SNR shows an elliptical shell-type structure at 1420 MHz, and has a 
408-1420 MHz TT-plot spectral index of $\alpha$=-0.5$\pm$0.1 (S$_{\nu}$$\propto$$\nu$$^{\alpha}$), typical of a shell-type SNR. The remnant's flux density
at 1420 MHz is 6.6$\pm$0.7 Jy, and at 408 MHz is 11.5$\pm$1.2 Jy. Both of these
are corrected for compact sources. An integrated spectral index of $-$0.45$\pm$0.13 is determined. This new SNR has among the lowest surface brightness of any 
known remnants ($\Sigma$$_{1 GHz}$=2.4$\times$10$^{-22}$ W m$^{-2}$ Hz$^{-1}$ 
sr$^{-1}$). 21 cm Stokes Q and U CGPS data (plus preliminary Effelsberg 
Q and U maps) show some suggestive features that correlate with total power. 
\ion{H}{i} observations show structures associated with G108.2$-$0.6 in the radial velocity range $-$53 to 
$-$58 km s$^{-1}$, and indicate it is located in the Perseus arm 
shock at a distance of 3.2$\pm$0.6 kpc. At this distance the diameter of 
G108.2$-$0.6 is 58 pc. 
IRAS maps (12, 25, 60 and 100$\mu$m) of the new SNR show rich infrared emission surrounding G108.2$-$0.6.

}
\titlerunning{Discovery of A New Faint Radio SNR G108.2$-$0.6}
\maketitle 

\section{Introduction}
There is discrepancy between the number of known Supernova Remnants (SNRs, 
about 265) and the number predicted by theory (about 10$^{3}$ Galactic SNRs, 
Li et al. 1991; $\sim$10$^{4}$ SNRs, Tammann et al. 1994). This has been 
considered the result of selection effects in current sensitivity-limited 
surveys, which favor the discovery of the brighter remnants (e.g. Green 2005; 
Tian \& Leahy 2004). Survey data from observations that combine high 
sensitivity and resolution with low radio frequencies are becoming the richest 
modern hunting-grounds for new SNRs. Some new SNRs have been discovered from the 
90 cm Very Large Array survey of the inner Galaxy (Brogan et al. 2006) and the 
Canadian Galactic Plane Survey (CGPS) of the outer Galaxy (Kothes et al. 2005). 
In this paper, we present the discovery of a faint shell-type SNR, through its
association with polarized radio emission and a steep-gradient radio spectrum.  

\section{Observations and Analysis}
The radio continuum, \ion{H}{i} and CO emission data sets come from the CGPS, 
which is described in detail by Taylor et al. (2003). These data are mainly 
based on observations from the Synthesis Telescope (ST) of the Dominion Radio 
Astrophysical Observatory (DRAO). The synthesized beam of the ST at 21 cm 
wavelengths is roughly 72$\arcsec \times$ 59$\arcsec$, and at 74 cm is 3.3
$\arcmin \times$ 2.8$\arcmin$. The velocity resolution in the 21 cm line is 
1.32 km s$^{-1}$, gridded to 0.82 km s$^{-1}$ per channel. DRAO ST observations 
are not sensitive to structures larger than an angular size scale of about 
3.3$\degr$ at 408 MHz and 56$\arcmin$ at 1420 MHz, thus for a complete sampling of 
spatial structures, the CGPS includes data from the 408 MHz all-sky survey of 
Haslam et al (1982), and the Effelsberg 1.4 GHz Galactic plane survey of Reich 
et al. (1997). Stokes U and Q data used here include all spatial-scales, by 
including observations from the DRAO interferometer, the Effelsberg Medium 
Latitude Survey (preliminary data, courtesy W. Reich and group at MPIfR), and by 
the DRAO 26-metre Survey (Wolleben et al. 2006). Short-spacing \ion{H}{i} line 
data are from the single-antenna 
survey of Higgs \& Tapping (2000), and have a resolution of 36$\arcmin$. IRAS 
data reprocessed for inclusion in the CGPS are also used for our study (Cao et al. 1997, Kerton $\&$ Martin 2000). 


\section{Results}
\begin{figure*}
\vspace{90mm}
\begin{picture}(80,80)
\put(-30,55){\includegraphics{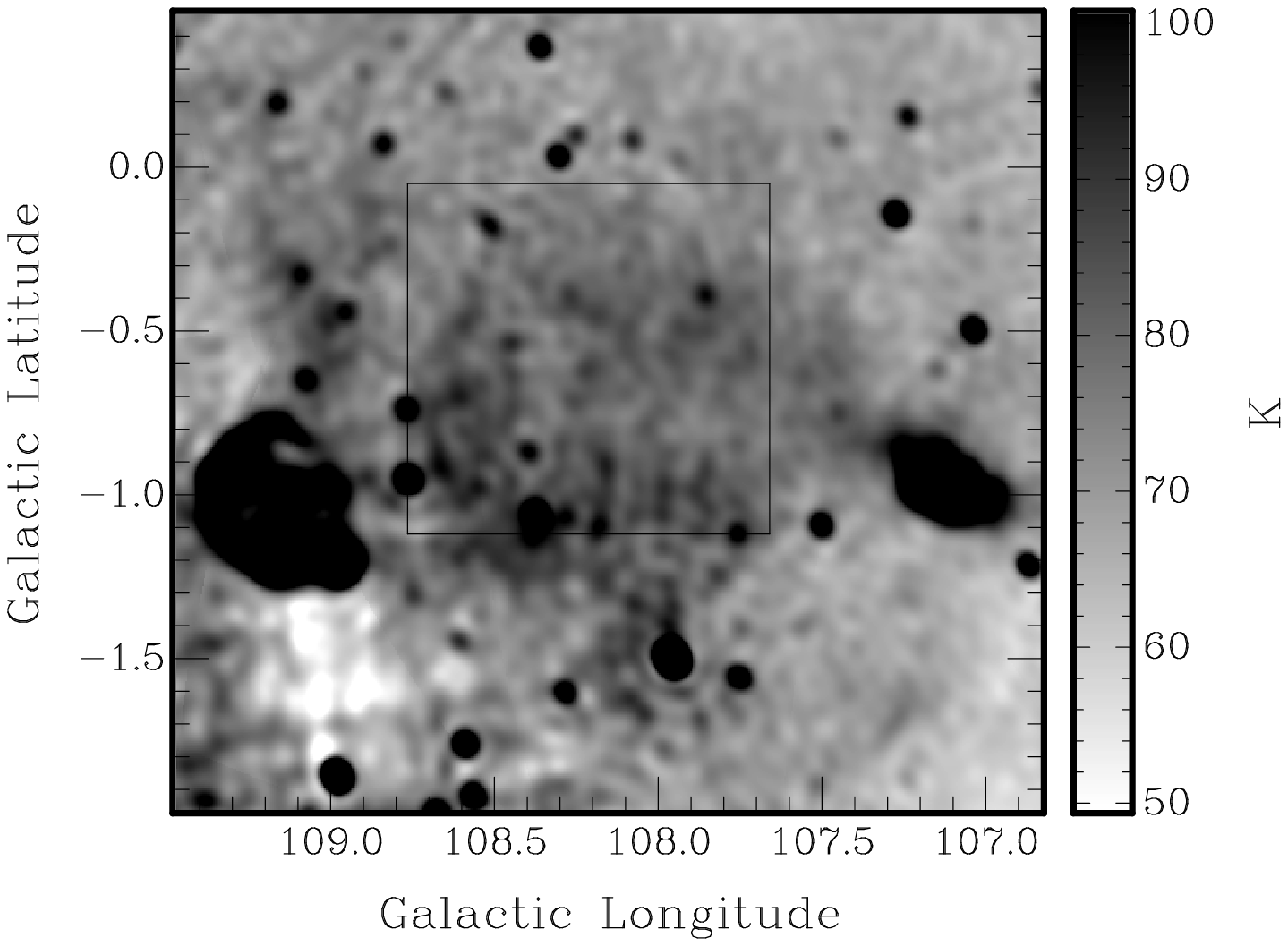}}
\put(240,55){\includegraphics{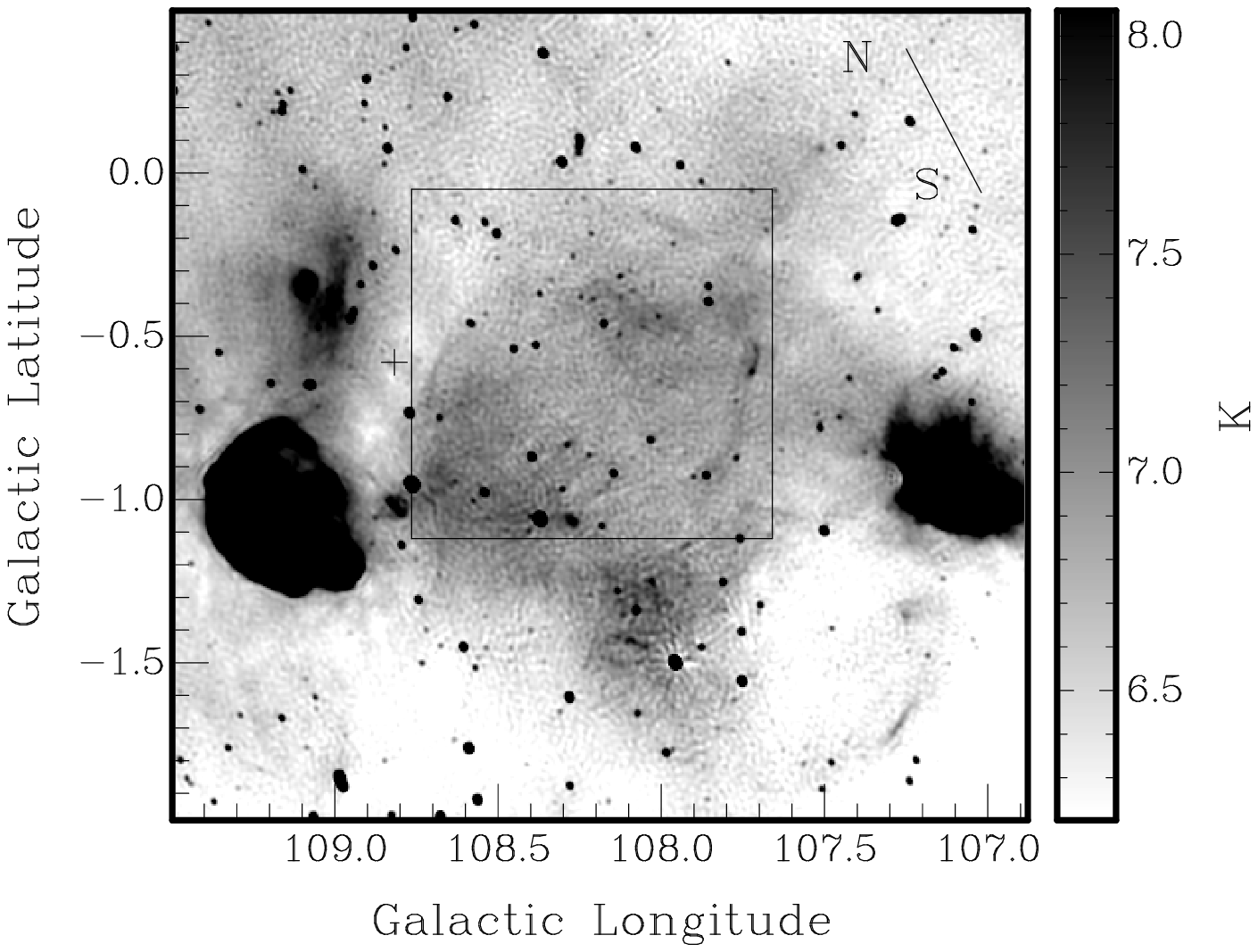}}
\put(-45,-10){\includegraphics{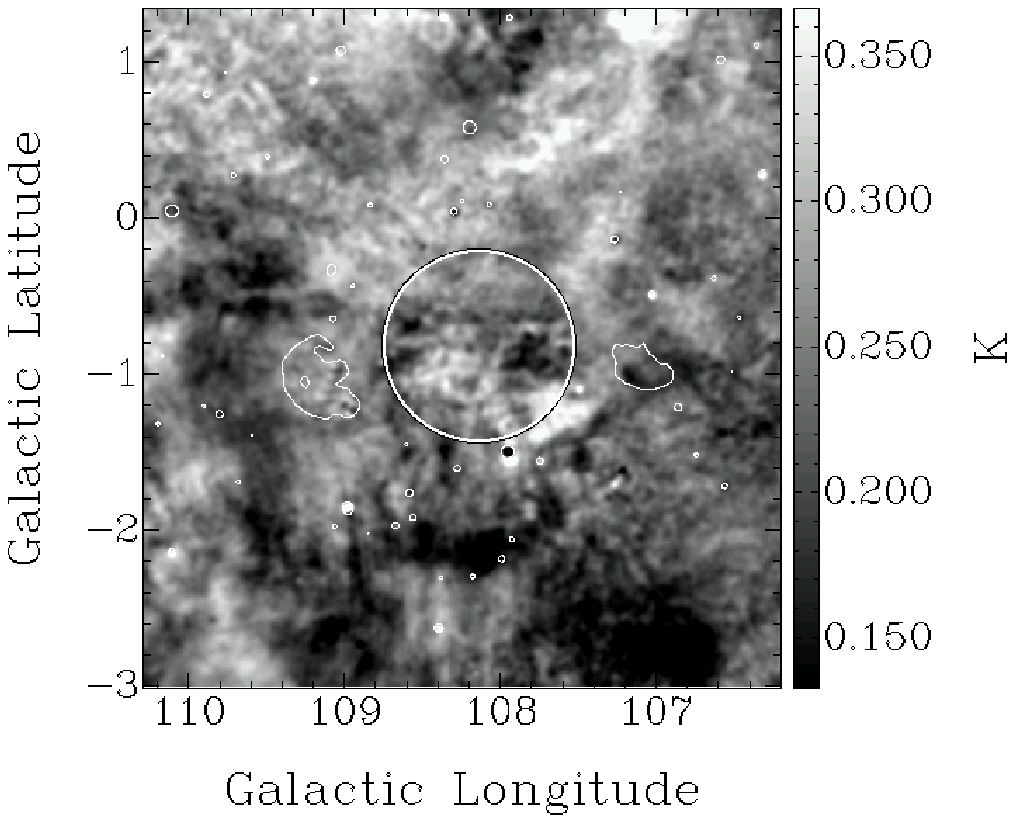}}
\put(225,-10){\includegraphics{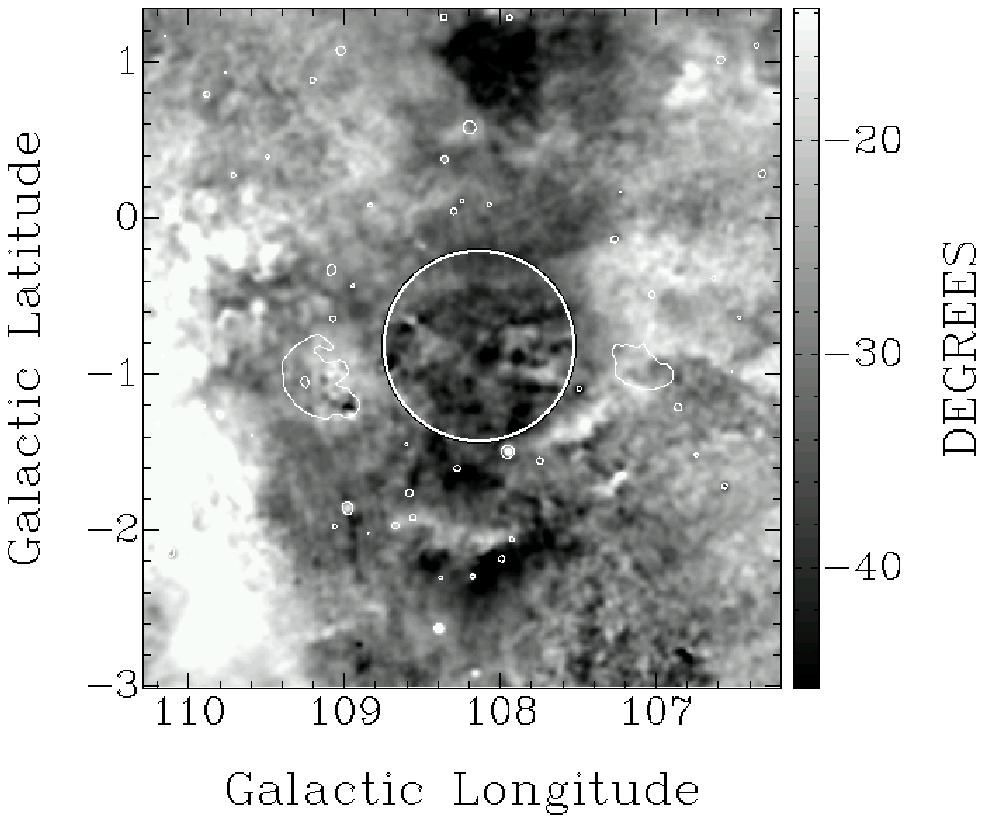}}

\end{picture}
\caption[xx]{The CGPS continumm maps at 408 MHz (left) and 1420 MHz (right) are shown in the first row. The box covering G108.2-0.6 is used for T-T plots spectral index. The position of pulsar B2255+58 and the direction of North (N) and South (S) are marked on the right plot.
 The second row presents the polarized appearance of G108.2$-$0.6, as seen at 21
cm in the CGPS. The polarized intensity map at left shows distinct emission (a small circular area; see text) from
within the continuum bounds of G108.2$-$0.6 (represented by the large circle,
as described in the text). The polarized-angle map at right shows that the
SNR lies within a distinct region of uniform, negative rotation angle. Both
maps have been smoothed to 3-arcminute resolution. Note the distinction
of G108.2$-$0.6 from the nearby CTB 109 (contour at 9 K; $\ell$=109.1$\degr$,~b=-1.0$\degr$),
which exhibits less polarized intensity, and a uniform rotation angle that
is more positive. Polarization data are from the CGPS and include preliminary data 
from the Effelsberg Medium Latitude Survey.}
\end{figure*}

\subsection{Radio Continuum Emission}
The new SNR has an average angular size 62$\arcmin$ (EW$\times$NS 70$\arcmin 
\times$ 54$\arcmin$, Fig. 1), and is located west of the nearby bright SNR CTB 109. South is another faint SNR G107.5-1.5 (Kothes 2003), and to the southwest is 
the bright radio \ion{H}{ii} region Sh2-142 ($\ell$=107.1$\degr$,~ b=-0.9$\degr$). Two bright compact \ion{H}{ii} regions (Sh2-146 \& Sh2-148) are 
immediately adjacent to the east and southeast of G108.2$-$0.6. The 
molecular cloud of Sh2-152 is found off the SE edge at 7 o'clock.

The SNR is characterized as a filled-centre region of diffuse emission (0.5 K 
above background at 1420 MHz), brightening slightly towards its perimeter to 
0.8 K above background near its northwest and southeast edges. The 1420 MHz map 
shows clear detail of the shell-type structure of G108.2$-$0.6. The shell's 
elliptical outline is nearly completely traced by filaments in the 1420 MHz 
map. The brightest filament of the SNR lies along its southwest boundary. A 
0$\degr$.6-radius circle centred on $\ell=$108.14$\degr,~b=-$0.81$\degr$ well 
delimits the object's nearly-circular appearance in radio continuum. Several 
compact sources within the SNR are resolved in both the 1420 MHz and the 408 
MHz maps. 

21 cm polarized emission is seen towards G108.2$-$0.6 (the lower plots of Fig. 1). The polarized 
emission is confined mainly to the low galactic latitude part of the remnant. 
A circular area of fairly well defined polarized intensity (PI, about 0$\degr$.4 in diameter) is centred on $\ell=$108.37$\degr,~b=-$0.975$\degr$. It is 
interesting to note that in maps of this area separated into 
thermal and non-thermal brightness temperature (made by assuming their radio 
spectra follow $\alpha=-$0.1 and $-$0.5, respectively), this patch spatially 
correlates with the brightest non-thermal emission from the SNR, which seems to 
form a crescent-like filament that wraps partially around the circular patch of 
PI (lower left of Fig. 1). The SNR's association with polarization is especially evident 
in a map of polarization angle (PA, the lower right of Fig. 1), which show the 
SNR's continuum shell encircles a region of roughly uniform angle ($-$39$\degr$$\pm$3$\degr$) over the SNRs face. The angle is uniformly negative with respect to 
the mean background angle (typically $-$28$\degr$ to $-$34$\degr$), and is also 
distinguished from the nearby CTB 109 ($-$23$\degr$, positive rotation with respect 
to the background). Polarized radio emission in SNRs originates with 
synchrotron emission (electrons interacting with the compressed ISM magnetic 
field), so some correlation of total power with polarized intensity and an 
ordering in angle makes sense (ignoring random depolarization and rotation by 
the intervening ISM). 

Notwithstanding the intriguing observed PI and PA features, it is difficult to 
firmly conclude that the polarized emission actually belongs to G108.2$-$0.6, 
as the larger region around it also shows random patches of PI of approximately 
the same relative brightness (though these patches are not correlated with any 
features in total power).

\subsection{Flux Density and Spectral Index}

The 1420 MHz CGPS map has many compact sources overlapping the face of 
G108.2$-$0.6. Bright compact sources affect the measured integrated flux 
densities for G108.2$-$0.6 and its measured spectral index. It is difficult to 
separate the SNR from diffuse background emission at 408 MHz due to 
both lower resolution of the observations and fainter emission of the new SNR at 
408 MHz. We can reliably correct for the effects of compact sources by 
subtracting them from the 1420 MHz map only (see Fig. 2), and 
then determining spectra by employing the special T-T plot spectral 
analysis method of Leahy (2006). The T-T plot method is described more 
generally in the paper of Tian \& Leahy (2005). For the T-T plot spectral index 
analysis, we select a single region covering the whole G108.2$-$0.6 shown in Fig. 1, which yields the T-T plot shown in the Fig. 3. Since compact sources are subtracted from the 1420 MHz map only, compact sources show up as a vertical line of points, with 408 MHz flux density but no 1420 MHz flux density.
We obtain a T-T plot spectral index of $\alpha$=-0.5$\pm$0.1 
(S$_{\nu}\propto \nu^{\alpha}$) between 408 MHz and 1420 MHz, a typical index for shell-type remnants. 

The integrated flux density of the new SNR is 6.6$\pm$0.7 Jy at 1420 MHz (ie $\Sigma$$_{1 GHz}$ = 2.4$\times$10$^{-22}$ Watt m$^{-2}$Hz$^{-1}$ sr$^{-1}$) and 11.5$\pm$1.2 Jy at 408 MHz, corrected for flux densities from compact sources within the SNR. Its flux density-based spectral index is $-$0.45$\pm$0.13 which is consistent with its T-T plot spectral index. We also check the 2695 MHz Effelsberg map of the region around G108.2-0.6, and find that G108.2-0.6 is marginally visible at this frequency (F\"urst et al. 1990). This supports the case of a steep spectrum. 
In terms of its radio surface brightness, this SNR is one of the 6 faintest known SNRs (Green, 2006). Two of them are in regions towards the anti-centre at high latitudes (G182.4+4.3, Kothes et al 1998; G156.2+5.7, Reich et al. 1992), three others are discovered by the CGPS survey in the inner parts of the Galaxy (G85.4+0.7 and G85.9+0.6, Kothes et al. 2001; G96.0+2.0, Kothes et al. 2005). Brogan et al. (2006) discovered 35 SNR candidates from the 90 cm VLA survey of the inner Galaxy (15 of 35 candidates are very confident), but their surface brightnesses are in the range of $\Sigma$$_{1 GHz}$=(1-15)$\times$10$^{-21}$ Watt m$^{-2}$Hz$^{-1}$ sr$^{-1}$. G108.2-0.6 is a large anti-centre object near the Galactic plane. Its discovery partly supports Kothes et al.(2001)'s conclusions that a survey with the characteristics of the CGPS can identify faint extended SNRs amongest the confusing thermal emission near the Galactic plane.

\begin{figure}
\vspace{40mm}
\begin{picture}(40,40)
\put(-5,-85){\includegraphics{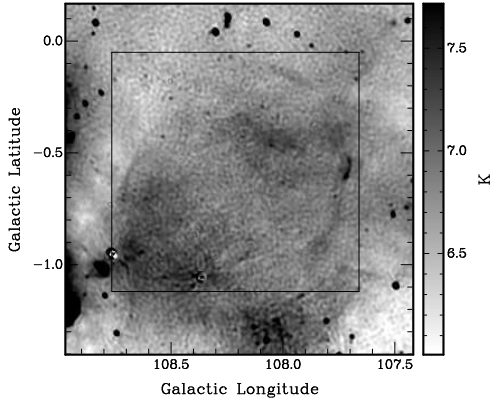}}
\end{picture}
\caption[xx]{The plot shows a zoomed 1420 MHz map with compact sources subtracted } 
\end{figure}

\begin{figure}
\vspace{50mm}
\begin{picture}(40,40)
\put(-50,195){\includegraphics{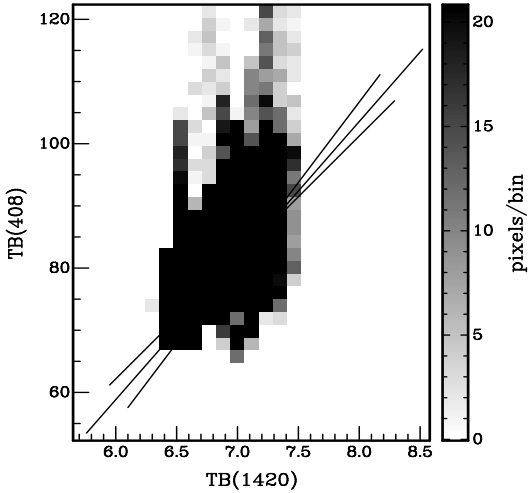}}
\end{picture}
\caption[xx]{408-1420 MHz T-T plot for the new SNR where compact sources have only been subtracted at 1420 MHz map for the plot. The fitted lines show the spectral slopes for $\alpha$=0.4, 0.5 and 0.6.}
\end{figure}

\begin{figure}
\vspace{40mm}
\begin{picture}(40,40)
\put(20,-55){\includegraphics{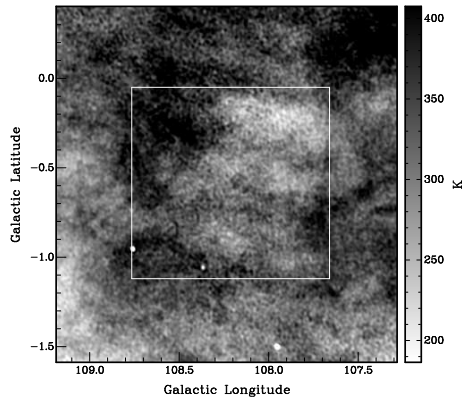}} 
\end{picture}
\caption[xx]{\ion{H}{i} emission in the field centered on G108.2$-$0.6 
integrated from $-$53 to $-$58 km s$^{-1}$. The box is the same as that in 
Fig. 1}
\end{figure}

\begin{figure}
\vspace{40mm}
\begin{picture}(40,40)
\put(-15,-100){\includegraphics{g108-iras60.eps}}
\end{picture}
\caption[xx]{IRAS mat at 60 microns in the field centered on G108.2$-$0.6. The box is the same as that in Fig. 1}
\end{figure}

\begin{figure}
\vspace{72mm}
\begin{picture}(0,65)
\put(0,0){\includegraphics{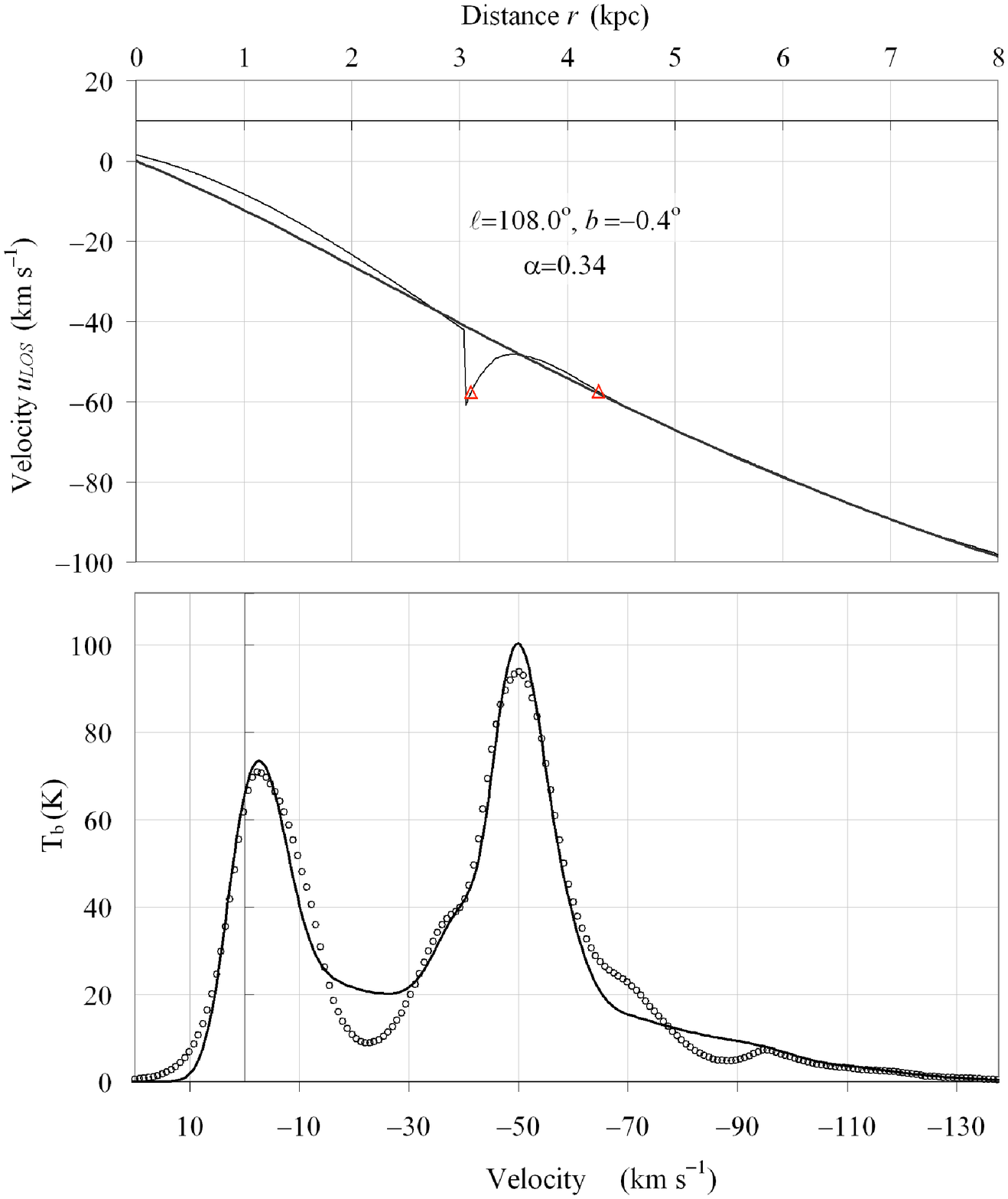}}
\end{picture}
\caption{The line-of-sight (LOS) velocity field towards the SNR (top
panel), as determined from the \ion{H}{i} modelling method of Foster \& 
MacWilliams (2006). The fit of the synthetic model spectrum to the observed 
one is shown at bottom.
The triangles in the velocity field mark the velocity and distance of G108.2$-$0.6
reported in this paper ($-$55 km s$^{-1}$); the ambiguous distance results
from the velocity reversal caused by the spiral shock (the front of which
is at 3.1 kpc). Density-wave streaming motions are terminated beyond the
Perseus arm, and no model component for the Outer Arm is included here (as
this LOS passes through very little of the far outer disk due to the
disk's upward warp). The LOS component of circular rotation only
(power-law of index $-$0.34; a gently-declining rotation curve) is shown as 
the smooth curve on the
velocity field. An spin-temperature of 200 K was assumed to make the \ion{H}{i}
disk optically-thin, and a turbulent dispersion width of 
$\sigma_{v}=$4 km s$^{-1}$ was used.}
\end{figure}

\subsection{HI, CO, Infrared, X-ray and Optical Emission}
We have searched the CGPS radial velocity range for features in the \ion{H}{i}
and $^{12}$CO lines ($^{12}$CO data from the Five College Radio Astronomy Observatory $^{12}$CO (J=1-0) outer galaxy survey, Heyer et al. 1998) which relate to the morphology of G108.2$-$0.6. We find
suggestive correlations with \ion{H}{i} features in the velocity range $-$53 
to $-$58 km s$^{-1}$. Fig. 4 shows map of \ion{H}{i} emission integrated over 
channels from $-$53 to $-$58 km s$^{-1}$. 
There are no apparent $^{12}$CO features associated with the remnant. We have 
checked IRAS 12, 25, 60 and 100$\mu$m maps of the region surrounding 
G108.2$-$0.6, and find rich diffuse infrared emission surrounding G108.2$-$0.6, particularly at 60$\mu$m (see Fig. 5). The main infrared emission is close to the eastern boundary and away from the SNR. Two bright regions of compact IR emission, \ion{H}{ii} regions 
Sh2-146 and Sh2-148, partly overlap on the east and southeast frontiers of 
G108.2$-$0.6, respectively. 

We have examined the ROSAT All-Sky Survey images (http://www.xray.mpe.mpg.de/cgi-bin/rosat/rosat-survey/), and find no obvious X-ray emission in the area of new SNR.  Optical images from the Palomar Digitized Sky Survey (http://skyview.gsfc.nasa.gov/cgi-bin/skvadvanced.pl) show no optical filamentary emission related with the new SNR.

\section{Discussion}
\subsection{The Distance}
With a radial velocity of $-$55$\pm$3 km s$^{-1}$ for the \ion{H}{i} structure 
around G108.2$-$0.6, and the assumption that this derives purely from circular 
galactic rotation, one estimates the SNR's kinematic distance to be $\sim$5.3 
kpc assuming circular velocity V$_{R}$=V$_{0}$=210 km s$^{-1}$ that is constant with 
Galactocentric distance, and R$_{0}$=7.6 kpc, as in Eisenhauer et al.(2005). 
However, as was shown by Foster \& MacWilliams (2006), the assumption of 
circular motion can lead to serious overestimation of a young object's 
distance, especially if that object is associated with the more enigmatic 
non-circular motions of the velocity field (like those introduced by the spiral 
shock, Roberts 1972). We therefore turn to the distance method of Foster \& 
MacWilliams (2006). The method fits a model of the density 
and velocity field to the observed \ion{H}{i} distribution, and calculates a 
kinematic distance from the fitted model. The model is not based on detailed 
hydrodynamical calculations, but rather is an analytical tool to functionally 
reproduce any reasonable geometric Galactic \ion{H}{i} density and velocity 
distribution through the disk. The model includes density and dynamics of the 
gas' linear response to a density wave (Lin et al. 1969), modified analytically 
to reproduce the non-linear behaviour of the velocity field (including the 
shock that precedes a spiral arm). As type Ib/II SNe are expected to occur near 
their progenitor's formation site (since the lifetime of massive stars is very 
short compared to their Galactic rotation periods), this distance method is 
particularly applicable to Galactic SNRs, which likely retain the kinematics of 
the spiral shock (presumably the place of formation of massive stars that give 
rise to them).

A range of fitted models in the line-of-sight $\ell$=108-109$\degr$,~b=-0.4-1.0$\degr$ 
(see Fig. 6) shows that the velocity range of $-$50 to $-$60 km s$^{-1}$ is 
covered by the Perseus arm spiral shock front. G108.2$-$0.6 is either just beyond this front (at $r$=3.2$\pm$0.6 kpc, or 
Galactocentric $R=$9.14 kpc) or has migrated through the Perseus arm ($R_{Per}$=
9.27 kpc) to a current position on its outer edge at $r$=4.4 kpc ($R=$9.93 
kpc). For a relative angular speed of $\left(\Omega-\Omega_{p}\right)=$1-2 km 
s$^{-1}$ kpc$^{-1}$ (pattern speed $\Omega_{p}$=20 km s$^{-1}$ kpc$^{-1}$), 
this migration would take on the order of 40-80 million years, longer than the 
lifetime of a massive progenitor which formed in the arm. Therefore, the nearer 
distance is preferred.

Since this SNR shares a similar velocity and distance with nearby CTB 109 
(3 kpc, Kothes et al. 2002), one presumes they may be related. However, CTB 109 is 
physically smaller than G108.2$-$0.6 (24 pc versus 58 pc respectively). Thus,
G108.2$-$0.6 is perhaps simply a more evolved remnant, or is evolving into a 
different medium. Its progenitor may have been part of the family of stars 
that formed in this complex area (which would include the progenitor of 
CTB 109). 

\subsection{G108.2$-$0.6 and PSR J2257+5909}
PSR J2257+5909 
is close to the northeastern edge of G108.2$-$0.6 (Fig. 1), has a large 
dispersion measure 
distance of 4.5 kpc from the Cordes $\&$ Lazio (2002) electron density model.
This distance is larger than that estimated for G108.2$-$0.6. Recent timing 
measurements (Zou et al. 2005) give PSR J2257+5909's proper motion 
550$\pm$120 km s$^{-1}$ ($\mu_{\alpha}$=24$\pm$6 mas yr$^{-1}$, $\mu_{\delta}$=-7$\pm$
5mas yr$^{-1}$). These show that PSR J2257+5909 is probably not associated 
with G108.2$-$0.6. 

\section{Conclusion}  
We have discovered a new SNR G108.2$-$0.6 using the data of the CGPS including 
Stokes I, Q, and U radio continuum, \ion{H}{i}, CO and infrared emission. The 
identification is based on its shell-type structure, a more-or-less typical 
non-thermal index of $\alpha=-$0.5. Based on a 
spatial correlation between \ion{H}{i} emission features and the SNR's 
continuum boundary, G108.2$-$0.6 is located in the Perseus arm shock at a 
distance of about 3.2 kpc. Its integrated flux density is 6.6$\pm$0.7 Jy at 
1420 MHz and 11.5$\pm$1.2 Jy at 408 MHz after subtraction of   
background and compact sources within the SNR. In X-ray and optical maps 
we find no significant emission from the SNR.
  
\begin{acknowledgements}
We would like to thank Wolfgang Reich and the polarization group at the 
Max-Planck-Institut f\"{u}r Radioastronomie (MPIfR) for allowing us access 
to preliminary Effelsberg Medium Latitude Survey polarization data. We 
acknowledge support from the Natural Sciences and Engineering Research Council of Canada. TWW appreciates support from the Natural Science Foundation of China. We thank Dr. R. Kothes at the DRAO for helpful discussion. 
The DRAO is operated as a national facility by the National Research Council of Canada.  The Canadian Galactic Plane Survey is a Canadian project with international partners. 
\end{acknowledgements}

\end{document}